\documentclass[prb,preprint,amsmath,amssymb]{revtex4}
\usepackage{siunitx}
\usepackage{graphicx}
\usepackage{dcolumn}
\usepackage{bm}
\usepackage{amssymb}
\usepackage{microtype}
\usepackage{xfrac}
\usepackage{color}
\usepackage{amsmath}
\usepackage{keyval}
\usepackage{natbib}
\usepackage{soul}
\usepackage{hyperref} 
\usepackage{xcolor}
\newcommand{\beginsupplement}{%
        \setcounter{table}{0}
        \renewcommand{\thetable}{S\arabic{table}}%
        \setcounter{figure}{0}
        \renewcommand{\thefigure}{S\arabic{figure}}%
     }

\DeclareGraphicsExtensions{.eps, .jpg,.pdf}

\begin{document}

\title{Anomalous metamagnetism in the low carrier density Kondo lattice YbRh$_{3}$Si$_{7}$}

\author{Binod~K.~Rai$^1$}
\author{S.~Chikara$^2$}
\author{Xiaxin~Ding$^2$}
\author{Iain~W.~H. Oswald$^3$}
\author{R.~Sch{\"o}nemann$^4$}
\author{V.~Loganathan$^1$}
\author{A.~M. Hallas$^1$}
\author{H. B.~Cao$^5$}
\author{M. Stavinoha$^6$}
\author{Haoran~Man$^1$}
\author{Scott~Carr$^1$}
\author{John~Singleton$^2$}
\author{Vivien~Zapf$^2$}
\author{Katherine~Benavides$^3$}
\author{Julia~Y.~Chan$^3$}
\author{Q.~R.~Zhang$^4$}
\author{D.~Rhodes$^4$}
\author{Y.~C.~Chiu$^4$}
\author{Luis~Balicas$^4$}
\author{A.~A.~Aczel$^5$}
\author{Q.~Huang$^7$}
\author{Jeffrey~W.~Lynn$^7$}
\author{J.~Gaudet$^8$}
\author{D. A. Sokolov$^9$}
\author{Pengcheng~Dai$^1$}
\author{Andriy~H.~Nevidomskyy$^1$}
\author{C.-L.~Huang$^1$}
\email[]{clh@rice.edu}
\author{E.~Morosan$^1$}

\affiliation{$^1$Department of Physics and Astronomy, Rice University, Houston, TX 77005, USA
\\$^2$National High Magnetic Field Laboratory, Materials Physics and Applications Division, Los Alamos National Laboratory, Los Alamos, NM 87545, USA
\\$^3$ Department of Chemistry, University of Texas at Dallas, Richardson, TX 75080, USA
\\$^4$ National High Magnetic Field Laboratory, Florida State University, Tallahassee-FL 32310, USA
\\$^5$ Neutron Scattering Division, Oak Ridge National Laboratory, Oak Ridge, TN 37831, USA
\\$^6$ Department of Chemistry, Rice University, Houston, TX 77005, USA
\\$^7$ NIST Center for Neutron Research, National Institute of Standards and Technology, Gaithersburg, MD 20899, USA
\\$^8$ Department of Physics and Astronomy, McMaster University, Hamilton, Ontario L8S 4M1, Canada
\\$^9$ Max Planck Institute for Chemical Physics of Solids, Dresden, 01187 Germany
}
\date{\today}

\begin{abstract}

We report complex metamagnetic transitions in single crystals of the new low carrier Kondo antiferromagnet YbRh$_3$Si$_7$. Electrical transport, magnetization, and specific heat measurements reveal antiferromagnetic order at $T_{\rm N} =$ 7.5~K. Neutron diffraction measurements show that the magnetic ground state of YbRh$_3$Si$_7$ is a collinear antiferromagnet where the moments are aligned in the $ab$ plane. With such an ordered state, no metamagnetic transitions are expected when a magnetic filed is applied along the $c$ axis. It is therefore surprising that high field magnetization, torque, and resistivity measurements with $H\| c$ reveal two metamagnetic transitions at $\mu_0 H_{1} =$ 6.7~T and $\mu_0 H_{2} =$ 21~T. When the field is tilted away from the $c$ axis, towards the $ab$ plane, both metamagnetic transitions are shifted to higher fields. The first metamagnetic transition leads to an abrupt increase in the electrical resistivity, while the second transition is accompanied by a dramatic reduction in the electrical resistivity. Thus, the magnetic and electronic degrees of freedom in YbRh$_3$Si$_7$ are strongly coupled. We discuss the orign of the anomalous metamagnetism and conclude that it is related to competition between crystal electric field anisotropy and anisotropic exchange interactions. \end{abstract}

\maketitle

\section{Introduction}

Materials containing partially-filled $f$ orbitals are of great interest to the strongly correlated electron system community because of their quantum complexity. This is driven by several competing parameters that include Kondo coupling (which favors a non-magnetic ground state with enhanced effective mass), Rudermann-Kittel-Kasuya-Yosida interactions (RKKY, which favors long-range magnetic order), and crystal electric field (CEF) effects (which left the degeneracy of the Hund's rule ground state multiplet). Among $f$-electron systems, the ground states of many Ce-, Yb- and U-based compounds are highly susceptible to tuning by non-thermal control parameters, such as pressure, chemical substitution, or magnetic field \cite{Loehneysen}. This often results in emergent phenomena such as unconventional superconductivity\cite{Schuberth and Steglich, Paglione and Maple, Mydosh and Oppeneer}, non-Fermi liquid behavior near a quantum critical point \cite{Gignoux,Custers and Coleman,Steppke and Brando,Palstra and Mydosh, Paglione and Maple,Binod}, hidden order \cite{Mydosh and Oppeneer} and metamagnetism\cite{Aoki1998,Prokes2001,Capan and Movshovich,Balicas and Fisk,Schmiedeshoff and Canfield,Morosan and Canfield, Lucas and Lohneysen, Deppe and Steglich, Sugiyama and Onuki}. In several materials, metamagnetism has been linked with magnetic quantum criticality\cite{Schmiedeshoff and Canfield,Grigera and Mackenzie1}, although the origin is unclear.

Interestingly, very few Yb-based metamagnetic (MM) compounds have been reported to date \cite{Tokiwa and Canfield,Matsubayashi and Uwatoko,Budko and Takabatake, Miyake2017}, compared to a larger number of Ce-based \cite{Lucas and Lohneysen, Deppe and Steglich, Capan and Movshovich, Aoki and Flouquet, Flouquet, Sugawara and Goto, Balicas and Fisk, Hirsoe and Onuki1,Tokiwa and Gegenwart} and U-based \cite{Aoki1998,Prokes2001} MM compounds. Here, we report the discovery of metamagnetism in single crystals of the Kondo lattice YbRh$_3$Si$_7$. YbRh$_3$Si$_7$ is the first compound displaying either Kondo correlations or metamagnetism in the ScRh$_3$Si$_7$ (1-3-7) family \cite{Chabot,Lorenz}, for which the only known rare earth-based systems are non-magnetic $R$Au$_3$Al$_7$ ($R$ = Ce-Sm, Gd-Lu) \cite{Latturner}, magnetic Eu(Rh,Ir)$_3$Ge$_7$ \cite{Falmbigl}, and YbAu$_3$Ga$_7$ with unknown physical properties \cite{Cordier}. In YbRh$_3$Si$_7$, the Kondo effect is clearly indicated by Kondo lattice-like resistivity, reduced magnetic entropy released at $T_{\rm N}$, and density functional theory (DFT) calculations. Anisotropic magnetic susceptibility and specific heat measurements reveal a long-range magnetic ordering transition at $T_{\rm N} = $7.5 K. Neutron diffraction measurements confirm that the zero field ordered state is antiferromagnetic (AFM), with the moments lying in the $ab$ plane.

We present high-field magnetization, torque, and resistivity up to 35 T. Around 2 K, these measurements reveal two field-induced MM transitions at $\mu_0 H_{1} =$ 6.7~T and $\mu_0 H_{2} =$ 21~T along the $c$ axis. Angular dependent magnetoresistivity shows that both $H_{1}$ and $H_{2}$ increase monotonically when the crystal is rotated away from the $c$ axis towards the $ab$ plane. When $H\perp c$, only one MM transition is found, up to the maximal field, at $\mu_0 H_{1} =$ 10~T. This behavior is starkly different from what has been observed in other MM materials; MM transitions are rarely observed for a field orthogonal to the moments, and, if present, typically occur at higher fields than those for the field parallel to the moments. Given that the easy axis, determined by the CEF anisotropy, is along the $c$ axis, while the ordered moment is lying in the $ab$ plane, the anomalous metamagnetism in YbRh$_3$Si$_7$ may be a result of the delicate balance among different underlying energy scales, including CEF anisotropy and exchange anisotropy. Understanding the metamagnetism in YbRh$_3$Si$_7$ will help draw a more complete picture of how subtle quantum effects steer the macroscopic behavior of different materials.
 
\section{METHODS}

Single crystals of YbRh$_3$Si$_7$ were grown from a Rh-Si excess liquid solution, using Yb (99.9999\%), Rh (99.95\%) and Si (99.99\%). The mixture was slowly cooled from 1200 $^{\circ}$C to 1100 $^{\circ}$C, and when the excess flux was decanted, rhombohedral crystals with typical dimensions  ~$3-5$ mm were obtained. The as-grown single crystals were subsequently annealed up to 200 hours under partial Ar atmosphere at 850 $^{\circ}$C. 

Room temperature powder x-ray diffraction patterns were collected in a Bruker D8 diffractometer using Cu K$\alpha$ radiation and the patterns were refined using the TOPAS software. Additional single crystal diffraction measurements were performed in a Bruker D8 Quest Kappa diffractometer equipped with an I$\mu$S microfocus source ($\lambda$ = 0.71073 \AA) operating at 50 kV and 1 mA, a HELIOS optics monochromator, and a CMOS detector. The crystal structure of YbRh${_3}$Si$_7$ was solved using direct methods in SHELXS2013 [Ref.~\onlinecite{Sheldrick1}] and all atomic sites were refined anisotropically using SHELXL2014 [Ref.~\onlinecite{Sheldrick2}] (see Table~\ref{TableI} and \ref{TableII} in Supplementary Materials). Powder neutron diffraction data were collected on the BT-1 high resolution neutron powder diffractometer and on the BT-7 triple axis diffractometer at the NIST Center for Neutron Research. We used an (002) pyrolytic graphite (PG) crystal as a monochromator with an incident beam energy of 14.7 meV and a PG filter to suppress higher order wavelength contaminations. Coarse collimations of Open-80’-80’ Radial collimations (FWHM) were employed, with a position-sensitive detector \cite{Lynn2012}. Single crystal neutron diffraction in zero magnetic field was performed on the four-circle diffractometer HB-3A at the High Flux Isotope Reactor (HFIR) at Oak Ridge National Laboratory (ORNL).  The data were collected at 4 K and 15 K by neutrons with a wavelength of 1.546 {\AA} from a bent perfect Si(220) monochromator \cite{Chakoumakos}. Single crystal neutron diffraction, with a magnetic field applied parallel to the $c$ axis, was performed on the fixed incident energy triple axis HB-1A at HFIR ($\lambda = 2.36$ \AA). A selection of Bragg peaks in the (HK0) scattering plane was measured at 1.8 K and 10 K in fields up to 8~T. The magnetic symmetry analysis was performed with SARAh \cite{Wills2000}. The nuclear and magnetic structure refinements were carried out with the FullProf Suite \cite{Rodriguez-Carvajal}.

Anisotropic temperature- and field-dependent DC magnetization measurements were done in a Quantum Design (QD) Magnetic Property Measurement System (MPMS) with a $^3$He option. Specific heat data were collected in a QD Physical Property Measurement System (PPMS) and a Dynacool with a dilution refrigerator using a thermal relaxation method for temperatures between 0.05 K and 30 K and magnetic fields up to 9 T. The temperature-dependent AC resistivity of bar-shaped crystals was collected in the QD PPMS, with $f$ = 17.77~Hz and current $i = 1$~mA parallel to the $c$ axis using a standard four-terminal configuration. 

To elucidate the electronic and magnetic properties of YbRh$_3$Si$_7$,  band structure calculations were performed using DFT with the linearized augmented plane-waves as a basis, as implemented in the WIEN2K code~\cite{Wien2k}. The generalized gradient approximation was used to account for the exchange and correlations~\cite{PBE}, and the polyhedron integration method was used to calculate the electronic density of states (DOS). The effect of Hubbard $U$ was incorporated within the DFT+U method~\cite{Anisimov93}.

The high-field magnetization measurements up to 35 T were performed using an extraction magnetometry technique in a capacitor-driven pulsed field magnet at the pulsed field facility at Los Alamos National Laboratory. The change in magnetization, $\Delta M$ with magnetic field $H$, was obtained by integrating the induced voltage with the sample inside a compensated coil and subtracting the integrated voltage recorded in a subsequent sample-out background measurement. The pulsed-field magnetization was calibrated using DC measurements obtained  in a QD MPMS up to 7 T. The applied pulsed magnetic field was determined by the induced voltage in a coil, calibrated using the de-Haas van Alphen oscillations of copper \cite{Goddard}. High-field torque magnetometry and resistivity measurements were carried out at NHMFL in Tallahassee. The torque was measured on a bar-shaped crystal through a CuBe cantilever beam, whose deflection was measured \emph{via} capacitive techniques, and the resistivity was measured \emph{via} a conventional lock-in AC technique in a 35 T resistive magnet in combination with a $^3$He cryostat. 

\begin{figure*}[ht!]
\includegraphics[width=1\columnwidth]{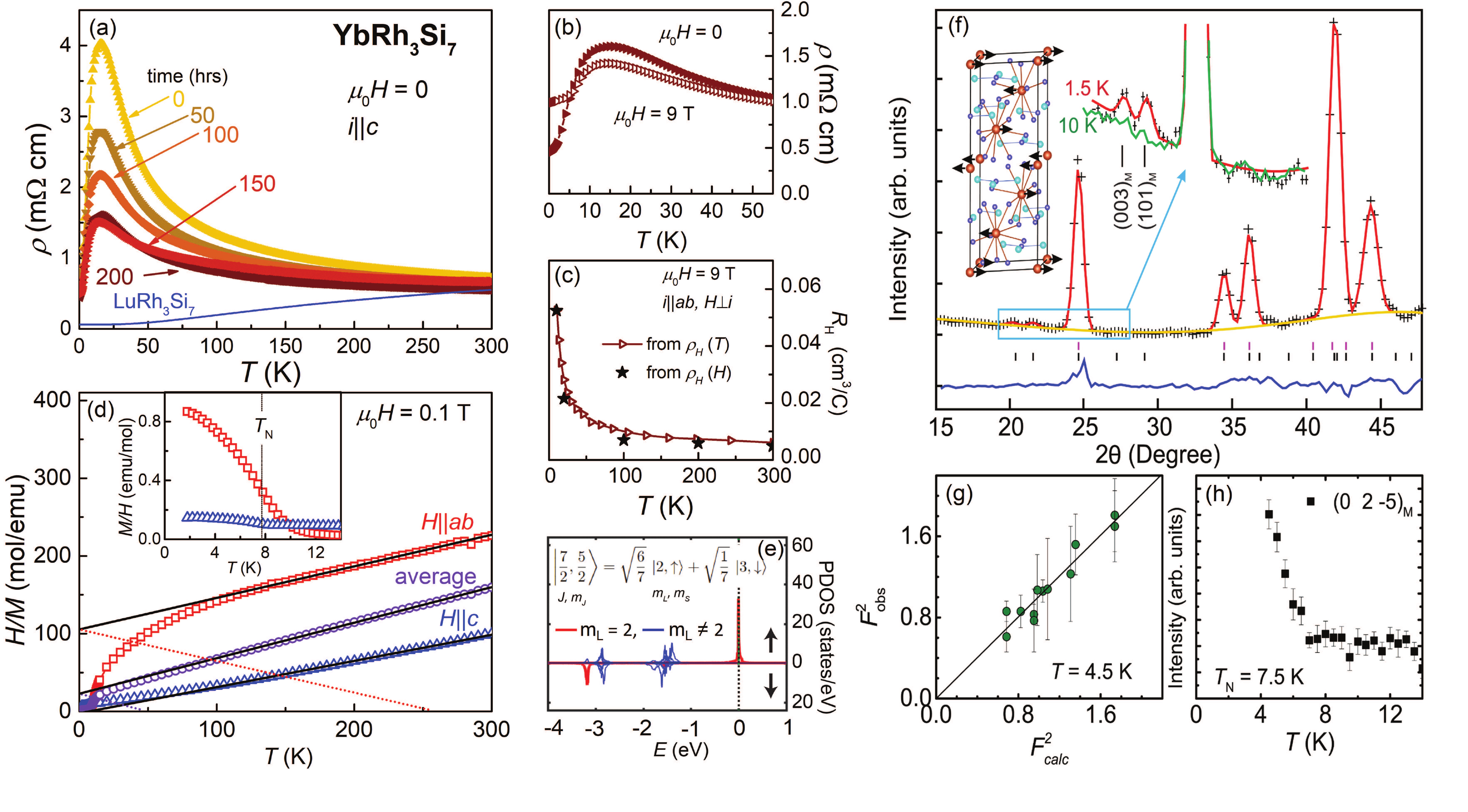}
\caption{\label{Rho} (a) Zero-field temperature-dependent electrical resistivity $\rho$ of YbRh$_3$Si$_7$ at different annealing times. The solid line is the data for LuRh$_3$Si$_7$. (b) Low-temperature $\rho$ at $\mu_0 H$ = 0 and 9 T applied parallel to the $c$ axis. (c) Hall coefficient R$_{\rm {H}}$ measured at $\mu_0 H$ = 9 T. (d) Inverse magnetic susceptibility $H/M$ vs. $T$ for YbRh${_3}$Si$_7$ with $\mu_0 H$ = 0.1~T parallel to $c$ (blue triangles) and  $ab$ (red squares) together with a polycrystalline average (purple circles) and Curie-Weiss fits (solid lines). Inset: low temperature magnetic susceptibility $M/H$. Note that 1 emu = 1 G cm$^3$ = 10 $^{-3}$ Am$^2$. (e) DOS projected onto the orbital angular momentum components of the Yb$^{3+}$ ion. (f) Powder neutron diffraction patterns at 1.5 K (data, crosses; Bragg peak positions, short vertical lines; difference between data and refined pattern, indigo curve), with the AFM structure of the system with collinear moments parallel to the $a$ axis (left inset) and an enhanced view of the (0 0 3) and (1 0 1) magnetic Bragg peaks at small angle (right inset). (g) Agreement of the single crystal magnetic structure refinement at $T$ = 4.5 K. (h) Magnetic order parameter measured at the (0 2 -5) peak upon warming. Error bars represent one standard deviation.} 
\end{figure*}
 
\section{Low carrier antiferromagnet with strong magnetic anisotropy}

YbRh$_3$Si$_7$ crystallizes in the ScRh$_3$Si$_7$ structure type with space group R$\bar{3}$c \cite{Chabot}. This structure features vertex-sharing YbRh${_6}$ octahedra and Si${_7}$ bipyramids oriented along the $c$ axis. The details of the crystal structure are presented in Supplementary Materials A. Because of the complexity of the rhombohedral R$\bar{3}$c crystal structure, it will be prudent to refer to the equivalent hexagonal structure and its respective $a$ and $c$ crystallographic directions. Likewise, when working in reciprocal space, all indices are given with respect to the hexagonal reciprocal lattice.

We first characterize the electrical transport properties of YbRh$_3$Si$_7$. The $H$ = 0 resistivity of the as-grown single crystals (t = 0, yellow symbols, Fig. \ref{Rho}(a)) shows semimetallic-like behavior where $\rho(T)$ increases upon cooling from room temperature down to $T^{\ast} \sim$ 20 K, a local maximum. By contrast, the non-magnetic analogue LuRh$_3$Si$_7$ is a normal metal (line, Fig.~\ref{Rho}(a)), albeit with a relatively large resistivity $\sim$ 1 m$\Omega$ cm at 300 K. The semimetallic $\rho(T)$ in YbRh$_3$Si$_7$ could either be intrinsic due to strong hybridization between the 4$f$ and conduction electron bands near the Fermi surface \cite{Dzero}, or extrinsic due to disorder-induced localization \cite{Anderson}, or both. Annealing effects at fixed temperature (850 $^{\circ}$C) with variable time (between 50 and 200 hours) allowed us to distinguish between these scenarios. The data in Fig. \ref{Rho}(a) (symbols) show a remarkable decrease of the absolute $\rho$ values over the entire measured temperature range as the annealing time increases, favoring a picture of reduced disorder with heat treatment. However, while the coherence maximum at $T^{\ast}$ is reduced by more than half from the as-grown to the optimally annealed samples (t = 150-200 hours, brown), the peak width and the semimetallic-like $\rho(T)$ behavior remain virtually unchanged. It can therefore be concluded that the electrical transport in YbRh$_3$Si$_7$ is an intrinsic characteristic of a Kondo lattice, where the negative resistivity coefficient (\textit{i.e.}, d$\rho$/d$T$ $<$ 0) at high temperatures is due to the spin-flip scattering of the conduction electrons off the magnetic centers. All other measurements reported here were performed on optimally annealed samples. The resistivity drop below $T^{\ast}$ is strongly field dependent (Fig.~\ref{Rho}(b)), indicating a magnetic phase transition. Magnetization, neutron diffraction, and specific heat measurements, which will be discussed below, show that YbRh$_3$Si$_7$ has an AFM ground state with $T_{\rm N}$ = 7.5~K. This is relatively high among Yb-based antiferromagnets, with very few other such systems showing comparable ordering temperatures (Yb$_3$Cu$_4$Ge$_4$ with $T_{\rm N} = 7.5$~K \cite{Dhar}, YbRhGe with $T_{\rm N} = 7$~K \cite{Katoh}, and Yb$_2$MgSi$_2$ with $T_{\rm N} = 9.5$~K \cite{Shah}).

Next we turn to the field dependent properties of YbRh$_3$Si$_7$, starting with the electrical resistivity and Hall effect. Large positive magnetoresistance at 2~K is apparent from the data measured at $\mu_0 H$ = 0 and 9~T (full and open symbols, Fig. \ref{Rho}(b)). The Hall coefficient R$_{\rm H}$ (Fig. \ref{Rho}(c)) is positive and strongly temperature-dependent. Within a single-band picture, the carrier density $n = 1/e\rm R_{\rm H}$ is estimated to be 1.5 $\times$ 10$^{21}$ cm$^{-3}$ at 2~K, one to two orders of the magnitude smaller than in a regular metal. We can therefore conclude that YbRh$_3$Si$_7$ is a low carrier Kondo lattice antiferromagnet. 

The $H =$ 0 magnetic ordering transition is confirmed by anisotropic magnetization measurements (Fig.~\ref{Rho}(d)). When the magnetic field is applied parallel or perpendicular to the $c$ axis, anisotropy is apparent in the magnetic susceptibility, with larger magnetization values along the $c$ axis, $M_{c}$, at high temperatures indicating axial CEF anisotropy. Above 100 K, Curie-Weiss behavior is evidenced by the linear inverse magnetic susceptibility $H/M$ (lines, Fig. \ref{Rho}(d)). A linear fit of the inverse average susceptibility $H/M_{ave}$, where $M_{ave} = (2M_{ab}+M_c)/3$, yields an effective moment $\mu_{eff}$ = 4.1 $\mu_{\rm B}$, close to the theoretical value for Yb$^{3+}$ ions $\mu_{eff}^{theory} = 4.54$ $\mu_{\rm B}$. Upon cooling below $\sim$ 10~K, a magnetic susceptibility crossover and a large (small) upturn in the $M_{ab}/H$ ($M_{c}/H$) [squares (triangles) in inset, Fig. \ref{Rho}(d)], resembling a ferromagnetic (FM) transition with moments perpendicular to the high-$T$ CEF axis. Ferromagnetic order along the hard axis has been observed in the heavy fermions Yb(Rh$_{0.73}$Co$_{0.27})_{2}$Si$_{2}$ \cite{Andrade} and YbNi$_4$P$_2$ \cite{Steppke and Brando}. The mechanism for this effect could be either quantum fluctuations or directionally-dependent transverse fluctuations \cite{Krueger}. However, in YbRh$_3$Si$_7$, neutron diffraction measurements reveal that the magnetic ground state is AFM. Thus, the $M/H$ crossover in YbRh$_3$Si$_7$ may have a different origin than in the above-mentioned hard axis ferromagnets.

To shed light on the magnetic properties of YbRh$_3$Si$_7$, electronic band structure calculations were performed using the DFT and DFT+U techniques, as described in the METHODS section. We have found the lowest energy configuration to be the one in which the Yb magnetic moments are ordered ferromagnetically within the $ab$ plane, pointing along the $c$ axis, with AFM order between adjacent planes. Next, we present the partial DOS projected onto different orbital angular momentum components, $m_L$ of the Yb$^{3+}$ ion, plotted in Fig.~\ref{Rho}(e). Top (bottom) panels show the minority (majority) spins, accordingly. Considering any given Yb ion, the minority spin DOS is dominated by $m_L=2$ (red) at the Fermi level (vertical dotted line). The other $m_L$ orbitals, which are represented by the blue lines, lie below the Fermi level  and hence do not contribute to the moment. This corresponds to the spin-orbit coupled state $|J=7/2, m_J=5/2\rangle$, as shown by the expression in the inset in Fig.~\ref{Rho}(e). We conclude that the ground state doublet is thus $|J=7/2, m_J=5/2\rangle$, and expect that, under large magnetic fields, this state will become fully polarized. Accordingly, the calculated saturated moment $\mu^{calc}_{sat} = (L_z + 2S_z)\mu_{\rm B}/\hbar $ should be 2.86 $\mu_{\rm B}$.

We next present neutron diffraction measurements, which allowed us to identify the zero field magnetic ground state of YbRh$_3$Si$_7$.  Upon cooling below $T_{\rm N} = 7.5$~K, these measurements reveal the formation of additional Bragg reflections as shown in Fig.~\ref{Rho}(f). These magnetic Bragg reflections were indexed with a $k = 0$ propagation vector in the R$\bar{3}$c space group. The best agreement with the measured diffraction pattern was obtained with the $\Gamma_5$ irreducible representation, which can be written as a linear combination of two basis vectors ($3\cdot\psi_5 + \psi_6$) \cite{Wills2000}. The resulting refinement is shown in Fig.~\ref{Rho}(f). In this collinear AFM structure, the spins are constrained to lie in the $ab$ plane, as represented for a single unit cell in the inset of Fig.~\ref{Rho}(f). The ferromagnetically-ordered planes are stacked antiparallel along the $c$ axis. This magnetic structure has also been verified with single  crystal neutron diffraction measurements. Figure~\ref{Rho}(g) presents the agreement between the measured and calculated structure factors. The temperature dependence of the (0,2,-5) magnetic Bragg peak, shown in Fig.~\ref{Rho}(h), reveals that at 4.5 K the intensity has not saturated. Thus, it is not surprising that the ordered moment determined from single crystal diffraction, 0.36 $\mu_{\rm B}$/Yb$^{3+}$ at 4.5 K, is slightly smaller than the moment derived from powder diffraction, 0.47 $\mu_{\rm B}$/Yb$^{3+}$ at 1.5 K. This partial order parameter gives $T_{\rm N}$ around 7.5 K, in agreement with the magnetic susceptibility measurements in the inset of Fig. \ref{Rho}(d). It is worth highlighting that the structure determined by neutron diffraction differs from the one used in the DFT calculations; while both structures have alternating AFM coupled planes, in DFT the moments point along the $c$-axis whereas experimentally they are found to point along the $a$-axis. This discrepancy is likely due to the fact that DFT may not properly account for the crystal electric field anisotropy.

\section{Kondo effect and hybridization}

\begin{figure}[hb!]
\includegraphics[width=1\columnwidth]{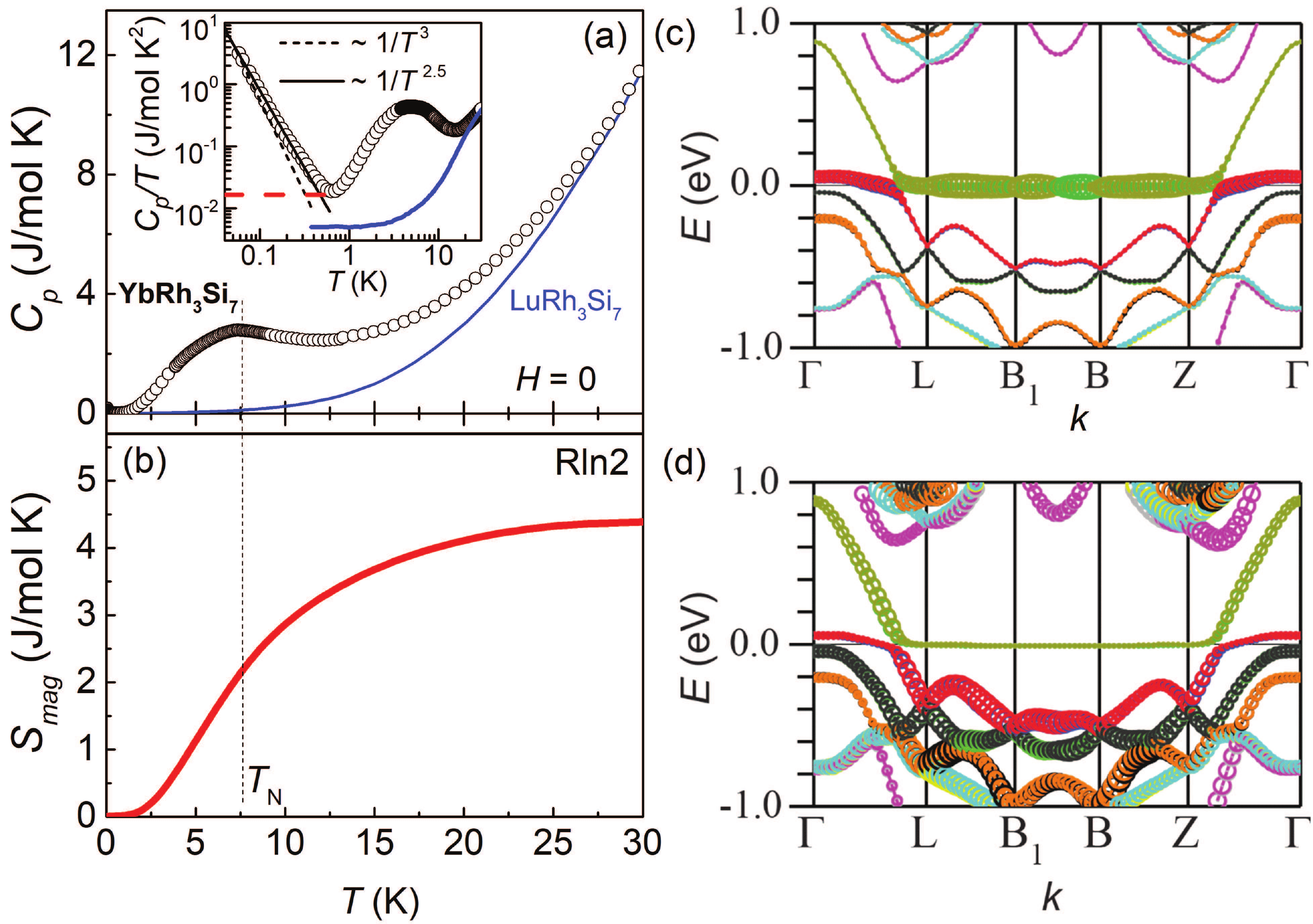}
\caption{\label{Cp} (a) $H$ = 0 $C_p(T)$ for YbRh${_3}$Si$_7$ (full circles) and its non-magnetic analogue (line) LuRh${_3}$Si$_7$. Inset: log-log plot of $C_p/T$ vs. $T$. (b) Magnetic entropy S$_{mag}(T)$. (c) Band structure of YbRh$_3$Si$_7$ with so-called ``fat bands" highlighting the contribution from Yb $f$ orbitals and (d) Rh and Si atoms. Thicker sections of the bands represent a larger partial contribution of the respective orbitals.}
\end{figure}

With the magnetic ground state resolved from neutron diffraction measurements, a better characterization of the correlations in YbRh$_3$Si$_7$ is needed, since the $H = 0$ resistivity, shown in Fig. \ref{Rho}(a), hinted at possible strong correlations and Kondo screening below $\sim$ 30 K. Long range AFM order at $T_{\rm N}$ = 7.5 K is marked by a broad peak in specific heat C$_p$ (Fig. \ref{Cp}(a)), consistent with the magnetic susceptibility and neutron data. For comparison, the specific heat data of the non-magnetic analogue LuRh$_3$Si$_7$ (solid line) is also shown in Fig. \ref{Cp}(a). The log-log plot of $C_p/T$ vs. $T$ in the inset reveals a power-law divergence of $C_p/T$ for YbRh$_3$Si$_7$ between 0.7 and 0.05 K. In this temperature range, such an increase of $C_p/T$ on cooling is often associated with an energy splitting of the nuclear quadrupole states in an electric field gradient of the individual atomic environment, \textit{i.e.}, $C_p/T \sim 1/T^{3}$. However, as shown in the inset of Fig. \ref{Cp}(a), the data  better described by a faster power law increase $C_p/T \sim 1/T^{2.5}$. This might indicate that the enhanced low $T$ specific heat is a convolution of a nuclear Schottky contribution and enhanced electronic specific heat contribution due to hybridization of the $f$ and conduction electron bands. A lower bound estimate of the magnetic entropy is obtained if we disregard the divergent contribution at the lowest temperatures. This lower bound is determined by assuming that below 0.7~K, $C_p/T$ is constant at  0.018 J/mol~K$^2$ (red dashed line in inset of Fig. \ref{Cp}(a)). Then we subtract the non-magnetic contribution as approximated by LuRh$_3$Si$_7$. The resulting magnetic entropy of YbRh$_3$Si$_7$ amounts to only 1/3 Rln2 at $T_{\rm N}$ (Fig. \ref{Cp}(b)), implying strong Kondo correlations, with a Kondo temperature around 15 K determined from $S_{mag}$(0.5 $T_{\rm K}) = 0.4$ Rln2. Even if we include the full divergent contribution at low temperatures, our entropy estimate only increases by a very small amount. Thus, the $S_{mag}$ in Fig. \ref{Cp}(b) can be viewed as slightly lower than the intrinsic value for YbRh$_3$Si$_7$, and a more accurate determination will rely on a precise estimate of the nuclear Schottky term, for which M{\"o}ssbauer and NMR measurements are necessary.

To elucidate the nature of the hybridization between Yb $f$ and conduction electron bands inferred from the thermodynamic measurements above, DFT+U calculations were performed \cite{Anisimov93} inside the AFM phase. The representative band structure is shown in Fig.~\ref{Cp} along the high-symmetry lines in the Brillouin zone. We have separated the partial contribution of the Yb $f$ electrons (Fig.~\ref{Cp}(c)) from that of the conduction electrons of Rh and Si (Fig.~\ref{Cp}(d)) using the ``fat band" representation, such that the thicker bands denote the larger contribution of the respective atomic orbitals. The results paint the canonical picture of hybridization between the very thin (atomic-like) Yb $f$-band and the parabolic conduction band, the latter with a bandwidth on the order of 1.5 eV. The chemical potential, pinned to the Yb $f$ level, crosses the conduction band along the $\Gamma$-Z direction, while the spectrum remains gapless (metallic in character) even in the AFM phase. The same observation is true for the paramagnetic (PM) phase (see Fig.~\ref{fatbands-PM} in Supplementary Materials~B).

An additional outcome from the band structure calculations is an estimate of the carrier density n $\sim~3.2 \times 10^{21}\,\text{cm}^{-3}$ in the PM phase, and $2.9 \times 10^{21}\,\text{cm}^{-3}$ in the AFM phase. These values are close to the experimental one, $n_\text{exp}\sim 1.5 \times 10^{21}\,\text{cm}^{-3}$ inferred from the low-temperature Hall coefficient (Fig.~\ref{Rho}(c)). This comparison is favorable, especially given that the experimental value was obtained within a simplified single-band model. Also, the DFT and DFT+U calculations are both based on a single-particle picture and, as such, do not capture the many-body Kondo-lattice phenomena. Nevertheless, the band structure calculations provide a reasonable starting point for the periodic Anderson model, and predict the system to be a low-carrier density Kondo metal, in agreement with the experimental results.

\section{Hard axis metamagnetism}

\begin{figure*}[ht!]
\includegraphics[clip,width=1\columnwidth]{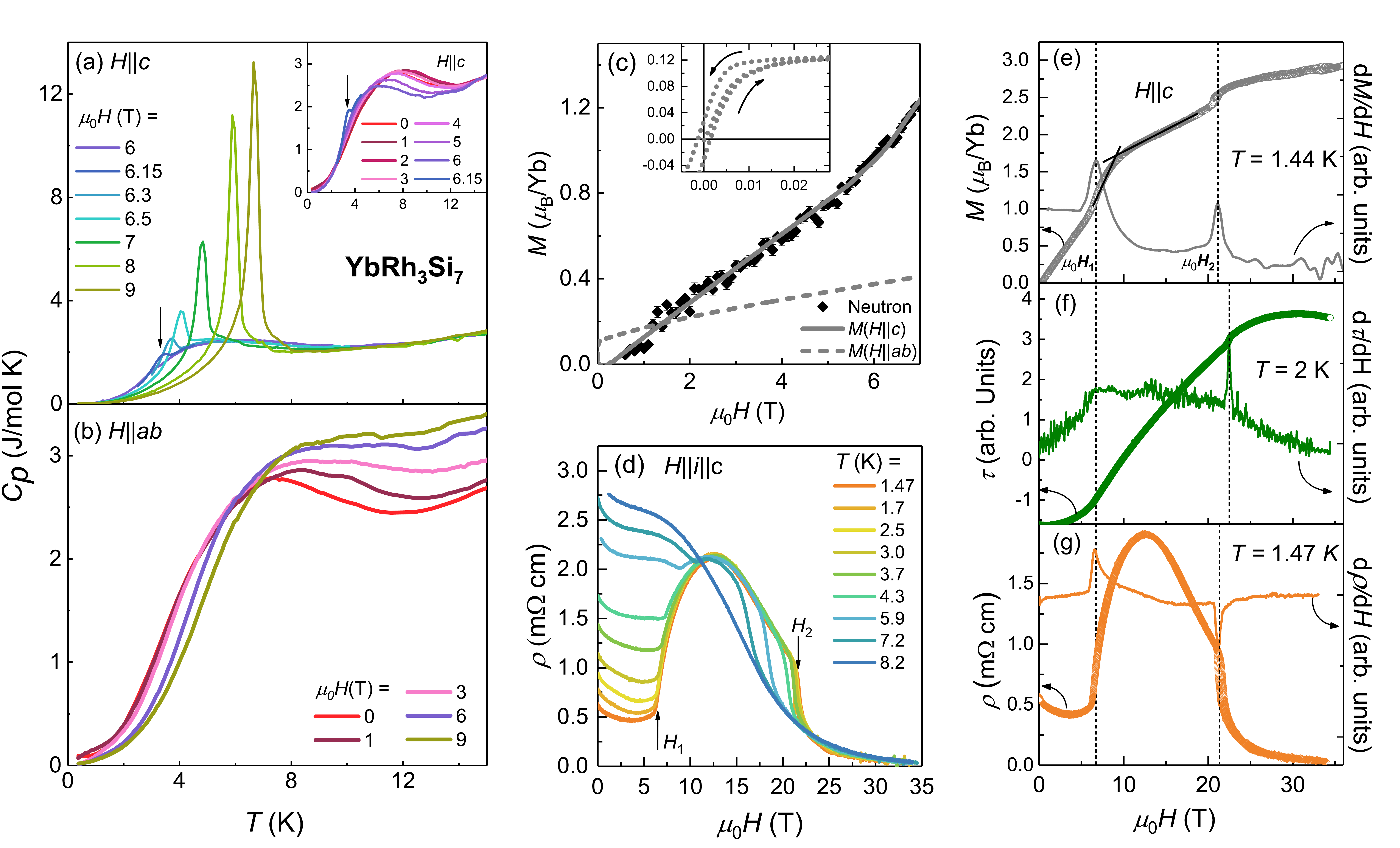}
\caption{\label{CpF} (a) $C_p$ vs. $T$ for YbRh${_3}$Si$_7$ with magnetic fields ($\mu_0 H$ $\geq{6~T}$) applied parallel to the $c$ axis. Inset: Low magnetic field data ($\mu_0 H\leq$ 6.15 T). (b) $C_p$ vs. $T$ with $H||ab$. (c) $M$ vs. $H$ at $T$ = 1.8 K for $H||ab$ (dashed line) and $H||c$ (solid line) up to $\mu_0 H = 7$ T. The $c$ axis moment from neutron diffraction is given by the solid symbols. Inset shows the low-field data for $H||ab$. (d) $H||c$ magnetoresistance isotherms. (e-g) Left axis: magnetization at $T$ = 1.44 K, torque at $T$ = 2 K, and resistivity at $T$ = 1.47 K as a function of magnetic field $H||c$. Right axis: derivative plots with respect to $H$. Dashed lines indicate the metamagnetic transitions at $\mu_0 H_1$ and $\mu_0 H_2$.}
\end{figure*}

Having provided evidence for the low carrier Kondo metal character in YbRh${_3}$Si$_7$, we now focus on how these properties are intertwined with the even more complex magneto-transport properties when the magnetic field is orthogonal to the $H = 0$ moment direction. In the $H\|c$ field-dependent specific heat, $T_{\rm N}$ appears to be suppressed slightly by fields up to $\sim$ 6 T (inset Fig. \ref{CpF}(a)). At higher fields, an additional transition becomes visible, and is marked by an arrow in Fig. \ref{CpF}(a) for $\mu_0 H$ = 6.15 T. As the field is further increased, the peak associated with this new transition becomes larger, sharper, and shifts to the higher temperature, reaching 6.8 K at $\mu_0 H = $ 9 T. In contrast, for $H\|ab$ (Fig. \ref{CpF}(b)), no additional transition is observed up to 9~T and and the $H = 0$ peak in $C_p$ monotonically shifts to higher temperatures. The behavior of $C_p$ with $H\|ab$ is reminiscent of the ``Vollhardt invariance'' that is associated with the transfer of specific heat weight under the constraint of magnetic entropy conservation. This effect is attributed to spin fluctuations above the ordering temperatures \cite{Vollhardt}. Similar behavior has been reported in several other strongly correlated systems, such as CeCu$_{5.5}$Au$_{0.5}$, MnSi, and CeAuSn \cite{Schlager,Janoschek,Huang}.

The magnetic susceptibility (Fig. \ref{Rho}(d)) and specific heat data (Fig. \ref{CpF}(a) and (b)) point to complex field-induced magnetic transitions and large CEF anisotropy in YbRh$_3$Si$_7$. Field-dependent thermodynamic and transport property measurements allow for an in-depth characterization of this complex magnetism. Low temperature magnetization measurements $M(H)$ up to 7 T not only confirm the magnetic anisotropy, but revealed a MM transition above $\mu_0 H = $ 6 T for $H\|c$ (solid line, Fig. \ref{CpF}(c)). In the orthogonal direction, no MM transition is observed (dashed line, Fig. \ref{CpF}(c)). The low field $M_{ab}$ shows a sharp increase with $H$ and a small hysteresis at very small fields (inset of Fig.~\ref{CpF}(c)). Due to the equivalence of the $a$ and $b$ hexagonal directions, the low field $M(H)$ behavior can be explained within a domain pinning picture because the spin configuration within the $ab$ plane is not unique. Indeed, this magnetic structure is composed of three symmetrically-equivalent domains, which can be generated from the configuration shown in Fig.~\ref{Rho}(f) by successively rotating the spins by \ang{120} in the $ab$ plane. Once a field is applied within the $ab$ plane, this symmetry is broken and the domain most closely aligned with the field will become energetically most favorable, resulting in its selection to form a single domain state. The very small $\sim0.01$~T barrier to this selection and the $\sim0.002$~T hysteresis are the result of domains being pinned by the low level of disorder that is present in any material.

To investigate the evolution of $M_c$ and temperature dependence of the MM transition at higher fields, we performed magnetization, torque, and magnetoresistance measurements up to 35 T as shown in Fig.~\ref{CpF}(d-g). All measurements were carried out for both increasing and decreasing  fields, and they indicate two MM transitions, around $\mu_0 H_1 \sim$ 6.7 T and $\mu_0 H_2 \sim$ 21 T. The values of $\mu_0 H_1$ and $\mu_0 H_2$ were determined from the peaks or dips in the derivatives, as shown in the right axes of Fig.~\ref{CpF} (e-g). The magnetization (Fig. \ref{CpF}(e)) varies linearly with $H$ up to the first critical field $\mu_0 H_1$. Following the MM transition at $\mu_0 H_1$, a linear increase of $M_c$ is seen for fields between 10 T and 21 T, followed by the second MM transition around $\mu_0 H_2$ = 21 T.  At higher fields, the magnetization  approaches a plateau close to 2.8 $\mu_{\rm B}$/Yb, much less than the Hund's rule ground state value of 4.5 $\mu_{\rm B}$ for Yb$^{3+}$. However, the measured value is consistent with the expected saturated magnetization for $m_j =$ 5$/$2, which was predicted by DFT calculations to be the CEF ground state (Fig. \ref{Rho}(e)). The critical fields $\mu_0 H_1$ and $\mu_0 H_2$ are also indicated by a change of slope in the magnetoresistance and torque data as shown in Fig.~\ref{CpF}(d,f-g). Temperature dependent magnetoresistance isotherms for $H\|c$ point to an increasing critical field value for $\mu_0 H_1$ and a decreasing critical field value for $\mu_0 H_2$ with increasing temperatures (Fig.~\ref{CpF}(d)). The two transitions converge around 12 T for $T\sim$ 8 K. 

\begin{figure}[tb!]
\includegraphics[clip,width=0.7\columnwidth]{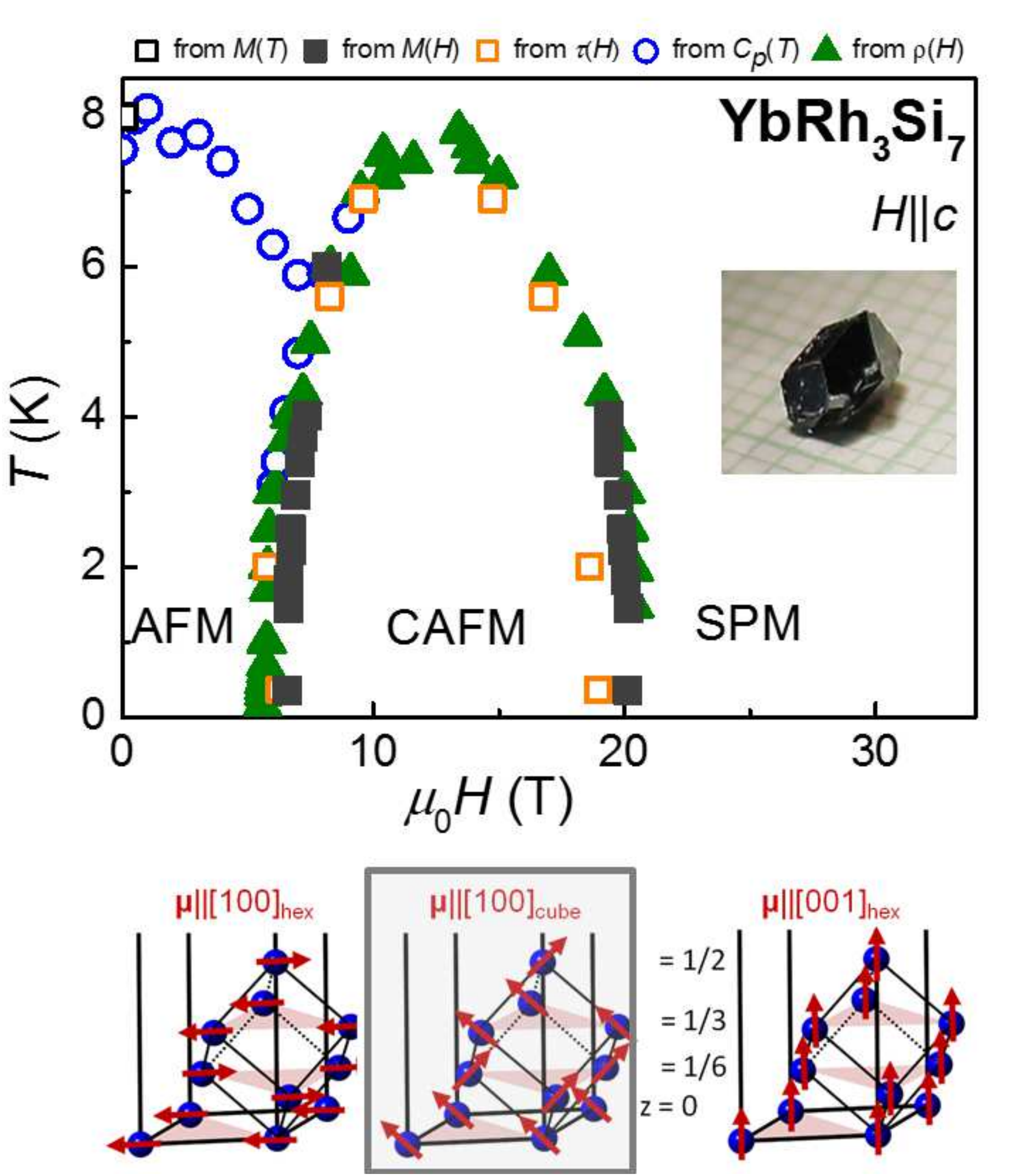}
\caption{\label{Phase} $T - H$ phase diagram of YbRh$_3$Si$_7$ together with three distinct magnetic spin configurations in different magnetic field regions. A picture of the crystal is shown in the inset.}
\end{figure}

\section{DISCUSSION}
YbRh$_3$Si$_7$ is a new low carrier Kondo metal, showing large CEF anisotropy at high temperatures and AFM order below $T_{\rm N}$ = 7.5~K. Zero field neutron diffraction measurements point to a collinear AFM magnetic structure, which evolves when a magnetic field is applied either parallel and perpendicular to the moments. Field-dependent magnetization measurements and theoretical calculations indicate that the phase above 20~T corresponds to the saturated $m_J = 5/2$ magnetic state. We therefore propose three distinct spin configurations to describe the different phases observed in the $H - T$ phase diagram for $H\|c$ in Fig. \ref{Phase}. In the low field regime, from zero field up to $\mu_0 H_1$, long range AFM order occurs, and this is schematically depicted in the bottom left plot in Fig.~\ref{Phase}. The magnetic moments $\mu$ (red arrows) are parallel to the $a$ axis $\mu \|[100]$, with FM $ab$ planes stacked antiferromagnetically along the $c$ axis. In the high field regime, $\mu_0 H \geq \mu_0 H_2$, all the moments are polarized parallel to the hexagonal $c$ axis $\mu \|[001]$, as indicated by both $M(H)$ measurements and DFT calculations. The corresponding spin-polarized paramagnetic (SPM) phase is illustrated in the bottom right plot in Fig. \ref{Phase}. Single crystal neutron diffraction measurements up to $\mu_0 H||c = 8$~T do not allow us to uniquely determine the magnetic structure in the intermediate phase. However, the $c$ axis ordered moment measured on the (110) Bragg peak is consistent with magnetization measurements, as shown in Fig.~\ref{CpF}(c). Above $\mu_0 H_{1}$, we do not observe the formation of any new magnetic Bragg peaks, as would be expected for an MM transition to a complex conical or helical state. Instead, we observe a sudden increase in the $k = 0$ FM moment. Thus, we propose a simple effective spin configuration for the intermediate phase that is consistent with our neutron diffraction measurements and can quantitatively describe the isothermal magnetization. The system enters a canted antiferromagnetic state (CAFM), with the moments aligned along the nearest-neighbour direction in the cubic Yb sublattice, which is at a 45 degree angle with the hexagonal axes (Fig. \ref{Phase}, bottom middle). 
The $H\|$[001] component of the magnetization for such a CAFM spin configuration can be calculated from $M_{001} = \mu$cos$\theta$ = 3$\times$(3.304\AA/5.468\AA) $\mu_{\rm B}$/Yb = 1.8 $\mu_{\rm B}$/Yb (schematically shown in Fig.~\ref{Cal} in Supplementary Materials~C). This value agrees well with the measured low $T$ magnetization around 9~T, $M_c$(9T;1.44K) $\sim$ 1.7 $\mu_{\rm B}/$Yb, indicated by the intersection of two violet lines (Fig. \ref{CpF}(e)).

The MM transitions in YbRh$_3$Si$_7$ are genuine phase transitions instead of crossovers, in contrast to some other metamagnetic systems \cite{Aoki and Flouquet,Tokiwa and Gegenwart,Miyake2017}. Similar phase diagrams with multiple MM transitions have been reported in other Ce- or Yb-based compounds, such as YbNiSi$_3$ \cite{Budko and Takabatake,Kobayashi}, YbAgGe \cite{Tokiwa,Fak}, CeAgBi$_2$ \cite{Thomas and Xia}, and CeAuSb$_2$ \cite{Marcus2018}. However, as expected, the MM transitions in these compounds occur for magnetic fields parallel to the $H =$ 0 moment direction ($M_0$). Remarkably, in YbRh$_3$Si$_7$ the MM transitions occur at lower fields for $H\perp M_0$. This is first apparent in the anisotropic $M(H)$ isotherms (Fig. \ref{CpF}(c)), with angular dependent magnetoresistance data reinforcing this point, as $H$ is rotated away from the $c$ axis towards the $ab$ plane (Fig.~\ref{angular} in Supplementary Materials~D). To our knowledge, YbRh$_3$Si$_7$ is therefore the only Ce- or Yb-based compound to show MM transitions for $H\perp M_0$, rendering the competition between the different energy scales (RKKY, CEF, Kondo etc) particularly complex. As DFT calculations already suggested, the energy scales for the moment parallel to $a$ and $c$ axes might be very close, and a modest magnetic field is sufficient to reorient the magnetic moment direction. This suggestion is supported by the isothermal magnetization data where a crossover of $M_{c}$ and $M_{ab}$ is observed around 1.6~T, as shown in Fig.~\ref{CpF}(c). When the applied field is larger than 1.6~T, the CEF easy axis anisotropy dominates and no crossover is expected, in contrast to the low field data shown in the inset of Fig.~\ref{Rho}(d). This is likely the result of field tuning of direct competition between the single ion CEF anisotropy and anisotropic exchange interactions, with the former dictating a magnetic easy axis parallel to the $c$ axis and the latter leading to the low field moment alignment in the $ab$ plane. However, this competition alone cannot explain the abrupt increase in resistivity that occurs at the onset of the MM transition at $\mu_0 H_1$ (Fig~\ref{CpF}(g)). In addition, in angle dependent magnetoresistivity measurements (Fig.~\ref{angular}), as the magnetic field is rotated away from the $c$ axis, a monotonic increase of $\mu_0 H_1$ and a monotonic decrease of resistance are observed, instead of extremes around $\theta =$ 45 degrees, i.e., the lowest energy state in the CAFM phase. There are some other possible underlying energy scales that are not discussed in this study including (1) a change in the Yb moment with field due to an excited CEF level crossing the ground state, such as a low-lying state below 1 meV, (2) a structural transition with field, or (3) a Kondo breakdown scenario where the itinerant 4$f$ electrons become localized at the MM transitions \cite{Kusminskiy}. The understanding of the exact nature of the anomalous metamagnetism in YbRh$_3$Si$_7$ warrants further study.   

In summary, YbRh$_3$Si$_7$ is a new low-carrier, antiferromagnetic, Kondo-lattice compound with anomalous metamagnetism. It serves as a forerunner among Ce- and Yb-based 1-3-7 analogues, rendering the 1-3-7 structure an ideal host structure to investigate the intertwinement of multiple energy scales including RKKY, Kondo, CEF anisotropy, and anisotropic exchange interactions.

\vspace{5mm}

\section*{Acknowledgements}
We thank P. Canfield, G. Lapertot, F. Steglich, C. Batista, L. Balents, and Q. Si for fruitful discussions. BKR, MS, CLH and EM acknowledge support from the Gordon and Betty Moore Foundation EPiQS initiative through grant GBMF4417. The work at University of Texas at Dallas was supported by NSF-DMR-1700030. The scientific work at LANL was funded by the LDRD program. This research used resources at the High Flux Isotope Reactor, a DOE Office of Science User Facility operated by the Oak Ridge National Laboratory. The NHMFL is funded by the U.S. NSF Cooperative Grant No. DMR1157490, the U.S. DOE, and the State of Florida. LB is supported by DOE-BES through award DE-SC0002613. PD acknowledge support from the U.S. DOE, BES DE-SC0012311 and the Robert A. Welch Foundation Grant No. C-1839. AHN and VL acknowledge the support of the Robert A. Welch Foundation Grant No. C-1818.  AHN was also supported by the U.S. NSF Grant No. DMR-1350237. Computational resources at Rice University were provided by the Big-Data Private-Cloud Research Cyberinfrastructure MRI-award funded by NSF under Grant No. CNS-1338099. The identification of any commercial product or trade name does not imply endorsement or recommendation by the National Institute of Standards and Technology.

\newpage
\section*{Supplementary Materials}
\beginsupplement

\renewcommand{\arraystretch}{0.5}
\setlength{\tabcolsep}{1pt}
\subsection{Crystallographic analysis}\label{App.cryst}

 \begin{figure}[h!]
\includegraphics[clip, width=0.7\columnwidth]{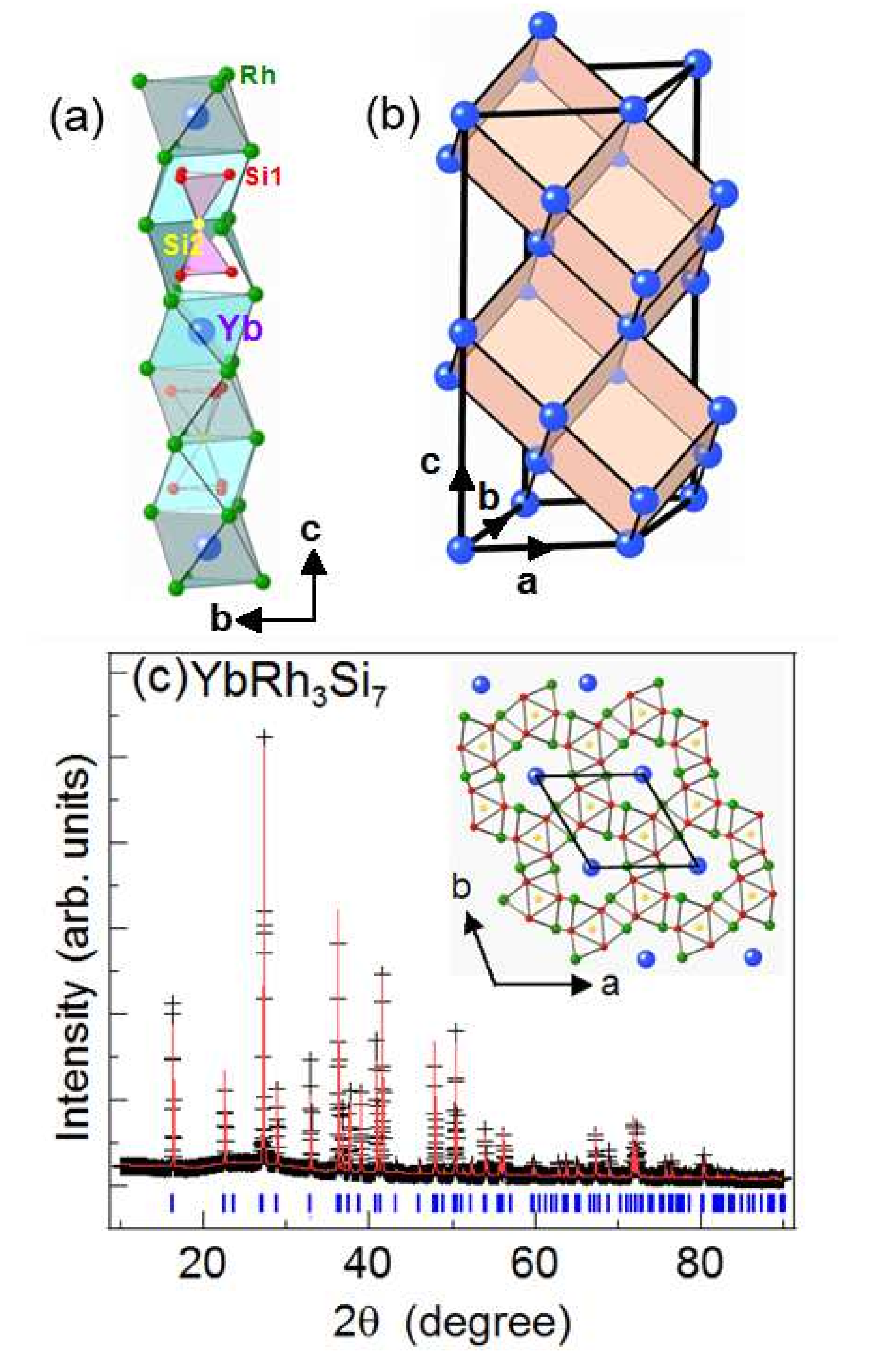}
\caption{\label{xrd} The crystal structure of YbRh$_3$Si$_7$ in the hexagonal lattice setting. (a) Face-sharing, nearly one-dimensional Rh octahedra (green) with corner-sharing Si double tetrahedra (pink) and Yb atoms (blue). (b) The Yb sublattice in one unit cell. (c) Room temperature powder x-ray diffraction pattern for YbRh$_3$Si$_7$ (black symbols) together with the calculated pattern (red line) for space group  R$\bar{3}$c and lattice parameters $a$ = 7.5482(4) {\AA} and $c$ = 19.8234(11) {\AA}. Inset: two-dimensional view of the crystal structure, viewed down the $c$ axis. All listed lattice parameters in the text and crystallographic directions correspond to the hexagonal-equivalent unit cell.}
\end{figure}

Table~\ref{TableI} shows the crystallographic data for as-grown and 150 hours annealed samples, with the atomic parameters of the latter given in Table~\ref{TableII}. 
\begin{table}[htbp]
\caption{\label{TableI} Crystallographic data for 150 hours annealed and as grown single crystals of YbRh$_3$Si$_7$ at $T$ = 298 K (space group R$\bar3$c)}
\begin{tabular}{c|c|c}
 \hline
	formula  									&  as grown	 		 		&  annealed (150 hrs)  \\  \hline   
	$a$ (\AA) 									&  7.5458(4)   				&  7.5482(4)  	 \\
	$c$ (\AA)									& 19.8240(11)   				&  19.8234(11)	 \\	$V$ (\AA$ ^{3}$) 								& 977.54(12) 				&  978.13(12)  \\

	absorption coefficient (mm$^{-1}$)					& 22.86					& 15.23	\\
	measured reflections							&7368						& 7697	\\
	independent reflections 							& 321 						& 319	  	\\
	R$_{int}$	 								& 0.063					& 0.039	\\
	goodness-of-fit on F$^2$	  						& 1.33					& 1.26		\\
	$R_1(F)$ for ${F^2}_o \textgreater 2\sigma ({F^2}_o)^a$	& 0.018					& 0.011		\\
	$wR_2({F^2}_o)^b$							& 0.046					& 0.023		\\
	extinction coefficient							& 0.054(4)					& 0.00245(17)	\\ \hline

  \end{tabular}
$^{a}R_1 = \sum\mid\mid F_o\mid - \mid F_c\mid \mid / \sum \mid F_o \mid~~~^bwR_2 = [\sum[w({F_o}^2 - {F_c}^2)^2]/ \sum[w({F_o}^2)^2]]^{1/2}$ 
\end{table}

\begin{center}
\renewcommand{\arraystretch}{1}
\setlength{\tabcolsep}{8pt}
\begin{table*}[h!]
\caption{\label{TableII} Atomic positions, site symmetry, $U_{eq}$ values, and occupancies for  150 hours annealed and as grown single crystals of YbRh$_3$Si$_7$.}
	\begin{tabular}{ c|c|c|c|c|c|c}
     \hline

Atom                       &Site symmetry      & x	          		& y	           	 & z	         		 & $U_{eq}$ (\AA$^2$)$^a$	    	 & Occupancy\\ \hline
as grown&  &  &  & 	&   &\\			
Yb	                    & $\bar{3}$ 		& 0			& 0			 &  0			& 0.00387(19)				& 1\\
Rh	                    & .2  			&0.32060(5)		& 0			 & $\frac{1}{4}$	& 0.0018(2) 					& 1\\
Si1	                    & 1 			&0.53733(13)	& 0.67865(14)	 & 0.02972(4)	& 0.0042(4)					& 1\\
Si2     	        	         & 3 2			&  0			& 0	 & $\frac{1}{4}$ 	& 0.0041(10)				& 1\\

annealed &  &  & 	&  & \\					
Yb	                    & $\bar{3}$ 		& 0			& 0			 &  0			& 0.00402(9)				& 1\\
Rh	                    & .2  			&0.32063(3)		& 0			 & $\frac{1}{4}$	& 0.00225(11) 				& 1\\
Si1	                    & 1 			&0.53739(10)	& 0.67873(10)	 & 0.02973(3)	& 0.0044(2)					& 1\\
Si2     	        	         & 3 2			&  0			& 0	 & $\frac{1}{4}$ 	& 0.0034(5)					& 1\\

    \hline

  \end{tabular}

$^a$ $U_{eq}$ is defined as one-third of the trace of the orthogonalized $U_{ij}$ tensor.

\end{table*}

\end{center}

\subsection{Details of the band theory calculations}\label{App.DFT}

\begin{figure}[h]
	\includegraphics[width=0.7\columnwidth]{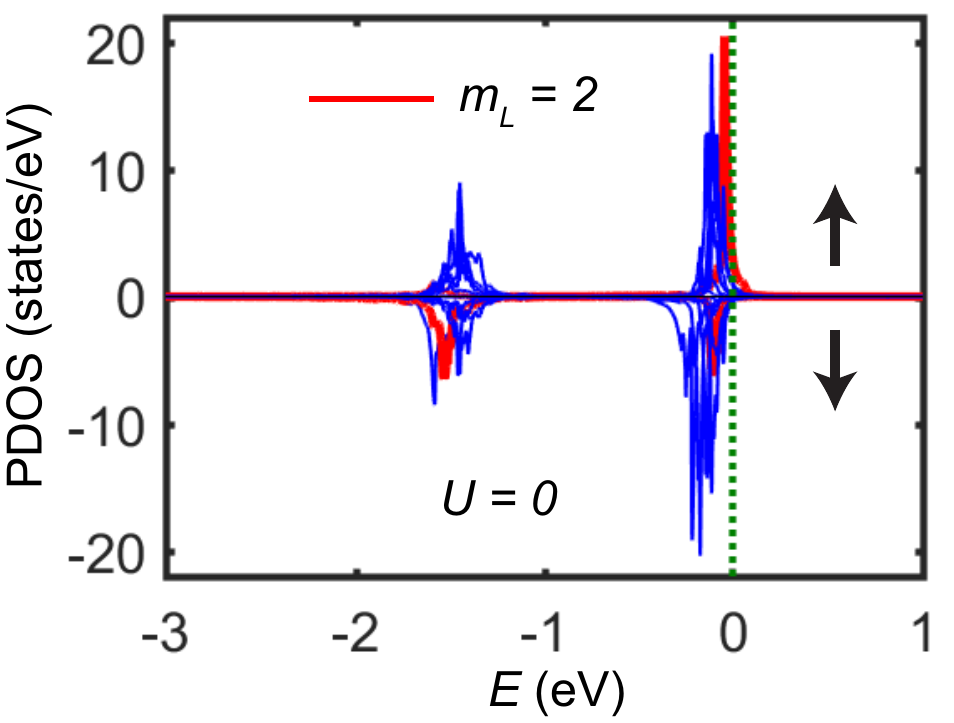}
	\centering
	\caption{DOS for the case without Hubbard interactions on Yb. The $m_L=2$ orbital (red) shows no clear split from the other orbitals. Top(bottom) panels show the minority(majority) spins. The vertical green dotted line represents the Fermi level.}
	\label{Fig:dosU0}
\end{figure}

In the main text, we showed that the DFT+U calculations in the AFM phase predict the $|J=7/2, m_J=5/2\rangle$ state as the ground state doublet of YbRh$_3$Si$_7$. For completeness, here we show the results obtained without including the Hubbard $U$ interaction.  The projected DOS plot in Fig.~\ref{Fig:dosU0} shows that the minority $m_L=2$ orbital (red line) has not fully split off from the other orbitals, unlike in the DFT+U case (Fig.~\ref{Rho}(e)). In this case, we find that the ordered moment is 0.75~$\mu_{\rm B}$/Yb. The inclusion of Hubbard interaction of $U=4$ eV on the Yb site further increases the moment to 1.5~$\mu_{\rm B}$/Yb, a characteristic of the DFT+U framework. Sufficient Hubbard interactions also cause the $m_L=2$ orbital to split from the other orbitals, as shown in Fig.~\ref{Rho}.

\begin{figure}
\includegraphics[width=0.7\columnwidth]{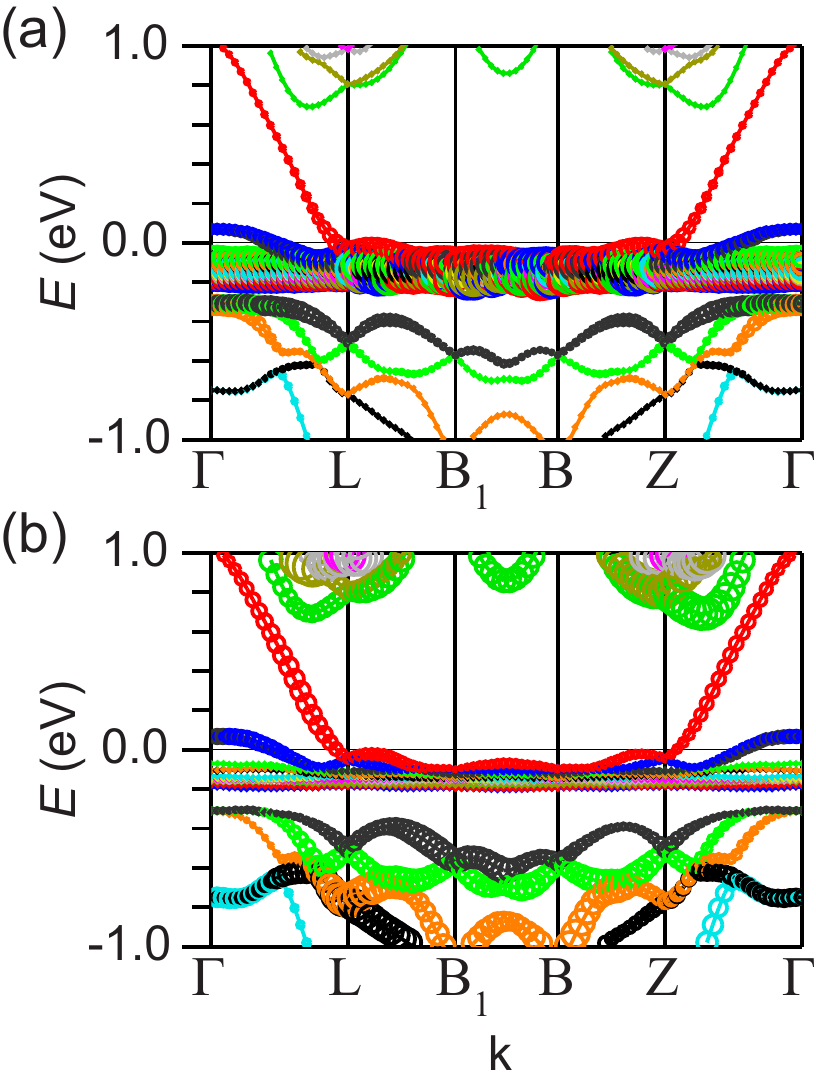}
\caption{\label{fatbands-PM} Band structure of YbRh$_3$Si$_7$ in the PM phase with so-called ``fat bands" highlighting the contribution from (a) Yb $f$-orbitals and (b) Rh and Si atoms. Thicker sections of the bands represent a larger partial contribution of the respective orbitals.}
\end{figure}

We note that both the DFT and DFT+U values of the ordered moment exceed the experimental value 0.36~$\mu_{\rm B}$/Yb obtained from the single crystal neutron diffraction. We attribute this disparity to the fact that the DFT-based single-particle framework cannot capture the many-body effects of the Kondo interaction and thus overestimates the bandwidths of Yb $f$-orbitals (and underestimates the effective mass). Additionally, the ordered moments are highly sensitive to the position of the Fermi level within the $f$-band.
 Despite this deficiency, the DFT+U calculations nevertheless  provide an adequate qualitative explanation of the magnetic properties, and in particular correctly predict the AFM structure and the the nature of the ground state doublet, together with the associated saturated moment (see Fig.~\ref{Rho} and the discussion in the main text).

Further information about the electronic properties in the PM state is obtained from the DFT band structure, shown in Fig.~\ref{fatbands-PM} where we have separated the partial contribution of Yb $f$-electrons (Fig.~\ref{fatbands-PM}(a)) from that of the conduction electrons of Rh and Si (Fig.~\ref{fatbands-PM}(b)). In this so-called ``fat band" representation,  the thicker bands denote the larger contribution of the respective atomic orbitals. This figure is to be compared with Fig.~\ref{Cp}(c-d) in the main text, the latter computed in the AFM phase using the DFT+U method. The two results are qualitatively the same in that they both show the hybridization between the Yb $f$-electrons and the conduction bands of Rh and Si. The difference is that in the present PM case, all 4 Kramers-degenerate $f$-bands corresponding to $J=7/2$ states appear close to the Fermi level and hybridize with the conduction electrons, whereas in the AFM case, the DFT+U calculation correctly selects out the $m_j=\pm 5/2$ state which appears at the Fermi level, as described above.

\subsection{Schematic view of the CAFM phase}\label{App.CAFM}
\begin{figure}
\includegraphics[width=0.7\columnwidth]{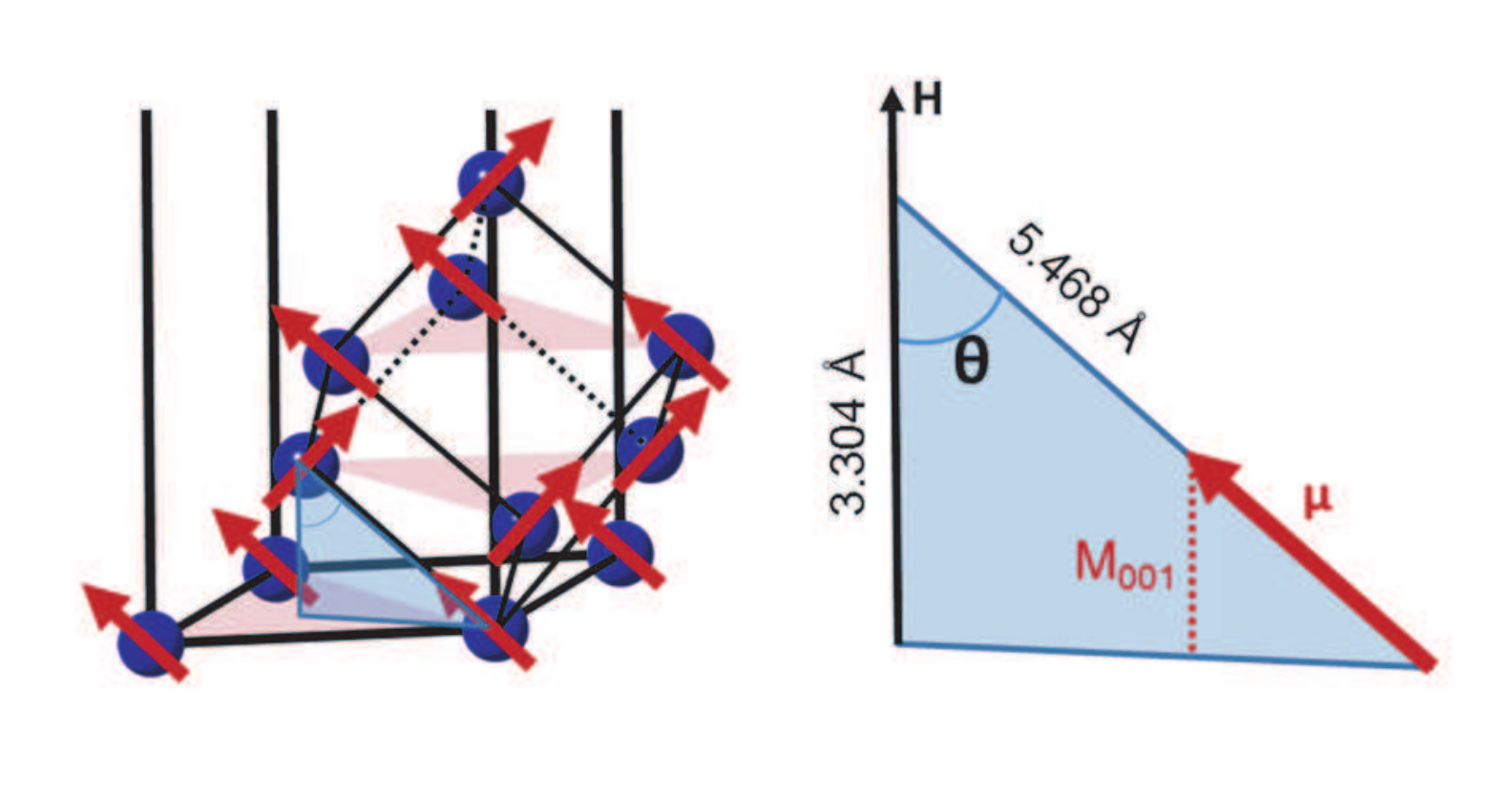}
\caption{\label{Cal} Magnetic spin configurations in the CAFM phase (left) and net magnetic moment ($M_{001}$) along the hexagonal $c$ axis.}
\end{figure}

\subsection{Angular dependent magnetoresistance}\label{App.Ang}
For angular dependent magnetoresistance measurements, the crystal was attached onto a sapphire chip with GE varnish and rotated from the $c$ axis towards the $ab$ plane. A Hall probe was used to calibrate the angle where the error bar was about a couple of degrees. The MM transition at $\mu_0 H_1$ moves up in field, from 6.7 T for $H\|c$ to 10 T for $H\|ab$, and $\mu_0 H_2$ moves above the available field range of 35 T for $\theta \sim$ 62 degrees.
\begin{figure}
\includegraphics[width=1\columnwidth]{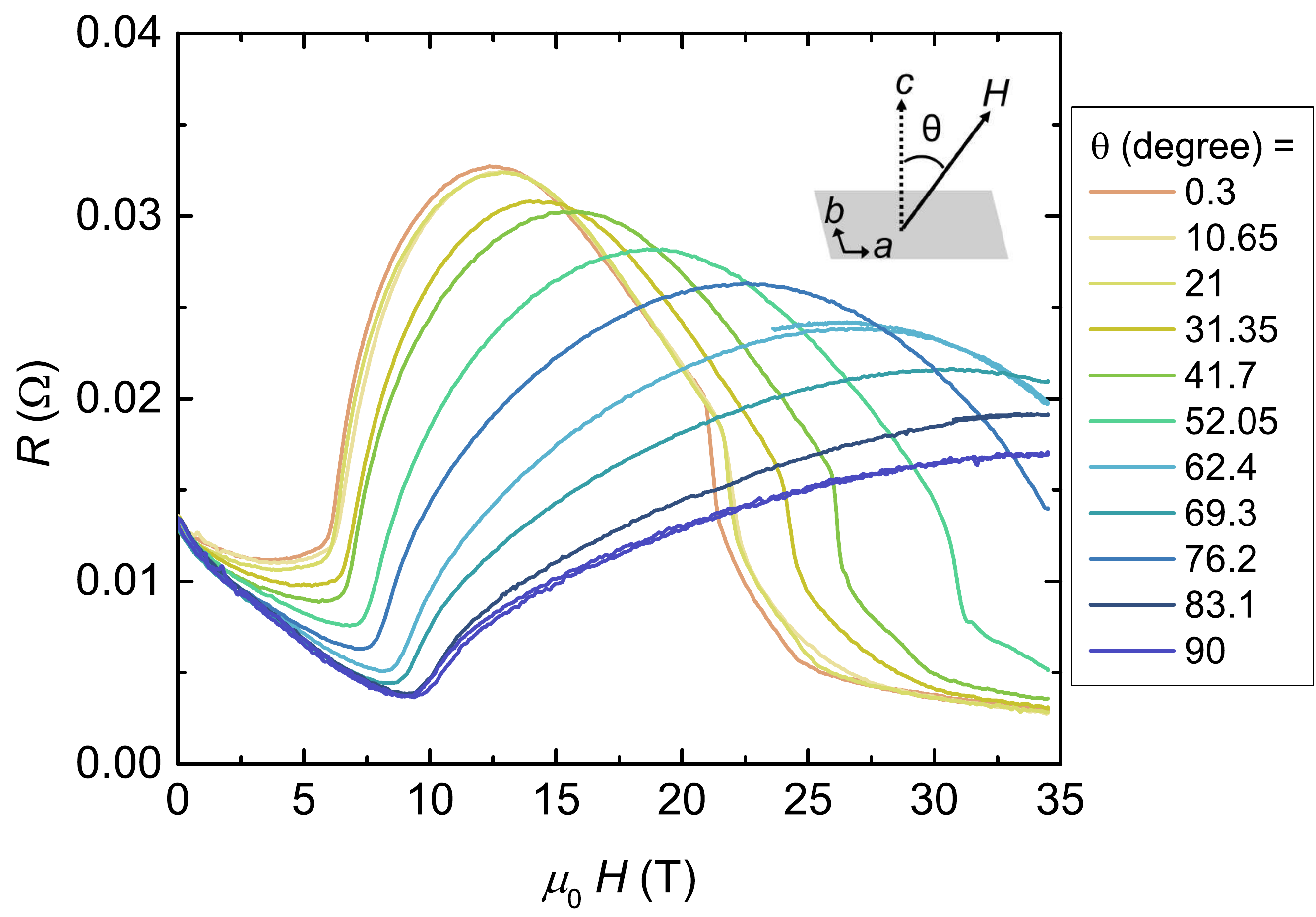}
\caption{\label{angular} Angular dependent magnetoresistance of YbRh$_3$Si$_7$ measured at $T =$ 1.47 K.}
\end{figure}

\end{document}